\documentstyle[11pt]{article}
\input epsf
\newcommand{\beq}[1]{ \begin{equation}\label{#1}}
\newcommand{\eeq}{\end{equation}}
\newcommand{\ba}{\begin{array}}

\newcommand{\beqa}{\begin{eqnarray}}
\newcommand{\eeqa}{\end{eqnarray}}
\newcommand{\bear}{\begin{array}{c}}
\newcommand{\bearr}{\begin{array}{cc}}
\newcommand{\ear}{\end{array}}

\newcommand{\mbf}{\mathbf}

\newcommand{\R}{I\kern-.3em{R}}
\begin{document}
\begin{titlepage}

\begin{flushright}
CPT-2002/P.4429\\
%CERN-TH YYYY
%hep-ph/00xxxx
\end{flushright}
\bigskip\bigskip

\begin{center}
\null
\vspace{1.5 cm}

{\Large \bf Impact-Picture Phenomenology for 
$\pi^{\pm}p,~ K^{\pm}p$ and  $pp,~ \bar p p $ \\
\vspace{0.5cm}
Elastic Scattering at High Energies}

\vspace{2.0cm}
\setcounter{footnote}{0}
\renewcommand{\thefootnote}{\arabic{footnote}}

{ \bf Claude Bourrely$^1$, Jacques Soffer$^1$ and Tai Tsun Wu$^2$ \\}
\bigskip
$^1$ Centre de Physique de Physique Th\'eorique\\
CNRS-Luminy case 907\\
13288 Marseille cedex 9, France\\

$^2$Gordon McKay Laboratory\\
Harvard University Cambridge, MA 02138, USA and\\
Theoretical Physics Division, CERN, 1211 Geneva 23, Switzerland

\vspace{2.0cm}
{\bf Abstract}
\end{center}
We present an extension to $\pi^{\pm} p$ and
$K^{\pm} p$ elastic scattering at high energies of the impact-picture
phenomenology we first proposed more than twenty years ago, for 
$p p$ and $\bar p p$ elastic scattering. We show, in particular, that the analytic form
of the opacity function for the proton obtained previously 
is compatible with the experimental results on $\pi p$ 
and $K p$ elastic scattering at high energies. It is proposed that $\pi^{\pm}$ and
$K^{\pm}$ external beams be provided from CERN-LHC, so that their elastic scattering
from protons can be studied at higher energies. Our 
phenomenology for $p p$ and $\bar p p$ elastic scattering is updated by
including new data and we give predictions for future experiments at BNL-RHIC and CERN-LHC. 

\end{titlepage}
%%%%%%%%%%%%%%%%%%%%%%%%%%%%%%%%%%%%%%%%%%%%%%%%%%%%
\section{Introduction}
\label{intro}
     In 1967, the Atomic Energy Commission, the predecessor to the 
present Department of Energy, announced the project to construct a 
200 GeV proton accelerator at the Fermi National Accelerator 
Laboratory.  This accelerator makes it possible to study the 
interaction between two protons both highly relativistic in the 
center-of-mass system.  This announcement motivated Cheng and one 
of the authors to initiate a program, using relativistic gauge 
quantum field theory, to study the high-energy behavior of hadron 
scattering.  Perhaps the most unexpected result from this study is 
that all total cross sections must increase without bound at high 
energies \cite{r1,r2,r3}.
     Another motivation for that investigation was to accept the 
challenge issued by Oppenheimer \cite{r4} at his concluding talk at the 
1958 Rochester meeting at CERN. He said at that time:

"There are areas where we know very little -- extremely high energy 
collisions, for example -- where little can be done by anyone."

     Three years after the theoretical prediction of the increasing 
total cross sections, experiments at the CERN Intersecting Storage 
Rings confirmed that the proton-proton total cross section indeed 
turns around and started to increase \cite{r5,r6}, and this data made  
possible to give the first quantitative predictions for future experiments
on the total cross sections for $ p\bar p, \pi^{\pm}p, K^{\pm}p$ \cite{r7}.
Careful phenomenology incorporating this increasing hadronic 
cross sections, called the impact-picture, was first carried out in 
1978 \cite{r8}.  Five years later, when more data became available, the 
parameters of this phenomenology were revised \cite{r9}.  It was found that 
the changes of the values are quite small.  These revised parameters 
have been used extensively to make various predictions, including for 
example the Coulomb interference effects and increasing integrated 
elastic cross section for proton-proton scattering up to 40 TeV in the 
center-of-mass energy, the proposed energy for the Superconducting 
Super Collider which was never built \cite{r10}.  All the predictions that 
can be confirmed experimentally have been confirmed \cite{r11}.  A recent chapter 
in a book gives a summary of the development up to this point \cite{r3}.
     Nevertheless, since it is by now nearly two decades since the 
1984 parameters were obtained, it is the opportune time to take another 
look at the parameters.  This is one of the purpose of the present 
paper.  Another purpose is to extend the phenomenology from the cases 
of proton-proton and proton-antiproton scattering to four other 
processes, namely $ \pi^{\pm}p, K^{\pm}p$.  The increasing total 
cross sections for these four processes were last investigated in 1973, 
nearly thirty years ago, and these predictions are in an even greater 
need of updating.  It should perhaps be mentioned that the 1973 
prediction for the $\pi^- p$ case has been well confirmed by later data \cite{r12}.

When we made our previous analysis in 1979 and 1984,
some experimental data were still preliminary, but since then they have been 
completed, so it is worth to "revisit" our model by making an analysis with a 
full set of final data points. In the next section, we recall
the main features of the model we use to describe elastic scattering
and we will explain our parametrization. Section 3 is devoted to the presentation
of the numerical results for $\pi p$, $K p$ scattering and for $p p$, $\bar p p$
scattering with some comparison between the present and the previous
determination of the parameters. We also give predictions for 
future experiments at BNL-RHIC and CERN-LHC. 
Our concluding remarks are given in section 4.

%%%%%%%%%%%%%%%%%%%%%%%%%%%%%%%%%%%%%%%%%%%%%%%%%%%%%%%%%%%%%%%%%%%%%%%
\section{\bf Description of the scattering amplitudes}
\label{des}
In the impact-picture representation, the
spin-independent scattering amplitude\footnote{ Here we neglect the spin-dependent
amplitude which was considered in
Refs.~\cite{r8,r13} for the description of polarizations and spin correlation parameters.}, 
for $p p$ and $\bar p p$ elastic scattering, reads as
\beq{ampli}
a(s,t) = {is \over 2\pi}\int e^{-i\mbf{q}\cdot\mbf{b}} (1 - 
e^{-\Omega_0(s,\mbf{b})})  d\mbf{b} \ ,
\eeq
where $\bf q$ is the momentum transfer ($t={-\bf q}^2$) and $\Omega_0(s,\mbf{b})$ is
defined to be the opaqueness at impact parameter $\bf b$ and at a given energy $s$. We take 
\beq{opac}
\Omega_0(s,\mbf{b}) = S_0(s)F(\mbf{b}^2)+ R_0(s,\mbf{b}) \ ,
\eeq
the first term is associated with the "Pomeron" exchange, which generates the diffractive
component of the scattering and the second term is the Regge background.
%$R_0(s,\mbf{b})$ is the Bessel transform of the Regge contributions.
The Pomeron energy dependence is given by the crossing symmetric expression \cite{r1,r2} 
\beq{energ}
S_0(s) = {s^c \over (\ln s)^{c'}} + {u^c \over (\ln u)^{c'}} \ ,
\eeq
where $u$ is the third Mandelstam variable.
The choice one makes for $F(\mbf{b}^2)$ is crucial and we take the Bessel transform of
\beq{formf}
\tilde F(t) = f[G(t)]^2 {a^2 + t \over a^2 -t} \ ,
\eeq
where $G(t)$ stands for the proton "nuclear form factor", parametrized like 
the electromagnetic form factor, as a two poles, 
\beq{fgt}
G(t) = {1 \over (1 - t/m_1^2)(1 - t/m_2^2)} \ .
\eeq
The slowly varying function occuring in Eq.(\ref{formf}), reflects the 
approximate proportionality between the charge density and the hadronic matter
distribution inside a proton.
So the Pomeron part of the amplitude depends on only {\it six} parameters
$c, c', m_1, m_2, 
f,$ and $a$. The asymptotic energy regime of hadronic interactions are
controlled by $c$ and $c'$, which will be kept, for all elastic reactions, at
the values obtained in 1984 \cite{r9}, namely
\beq{cc'}
c=0.167 ~~~~~~~~~~\mbox{and}~~~~~~~~~~~~c'=0.748~~.
\eeq
The remaing four parameters are related, more specifically to the reaction $pp$ ($\bar p p$)
and they will be slightly re-adjusted from the use of a new set of data.

We now turn to the Regge background. A generic Regge exchange amplitude has an expression
of the form
\beq{ampreg}
\tilde R_i(s,t)=C_ie^{b_it} \left[ 1 \pm e^{-i\pi\alpha_i(t)}\right](\frac{s}{s_0})^{\alpha_i(t)} \ ,
\eeq
where $C_ie^{b_it}$ is the Regge residue,
$\pm$ is the signature factor, $\alpha_i(t) = \alpha_{0i} + \alpha_i^{'} t $ is 
a standard linear Regge trajectory and $s_0 =1$GeV$^2$.
If we consider the sum over all the allowed Regge trajectories $\tilde R_0(s,t)= \sum_i \tilde R_i(s,t)$, the
Regge background $R_0(s,\mbf{b})$ in Eq. (\ref{opac}) is the Bessel transform of $\tilde R_0(s,t)$.
In $pp$ ($\bar p p$) elastic scattering,
the allowed Regge exchanges are $A_2$, $\rho$, $\omega$, so
the Regge background involves several
additional parameters, we will come back to them.

Let us now consider the case of $\pi p$ elastic scattering. The scattering amplitude
is also defined by Eq. (\ref{ampli}) and we keep the same structure for the opaqueness Eq. (\ref{opac}).
However, for the Pomeron exchange, we assume that the nuclear 
matter density is the product of the proton and the pion contributions. So with this reasonable
hypothesis, we write similarly to Eq.~(\ref{formf}), 
\beq{formfp}
\tilde F(t) =  f_{\pi} G(t) F_{\pi}(t) {a_{\pi}^2 + 
t \over  a_{\pi}^2 -t} ~.
\eeq 
Here $G(t)$ is given by Eq.~(\ref{fgt}) with the same parameters values $m_1,~m_2$ as
in the proton case and $F_{\pi}(t)$ is the pion "nuclear form factor", we parametrize
like the pion electromagnetic form factor, as a single pole $1/(1 - t/m_{3\pi}^2)$ \cite{BEB78}. 
Therefore the Pomeron term for $\pi p$ elastic scattering involves
{\it three} additional free parameters $ f_{\pi}$, $m_{3\pi}$, and $ a_{\pi}$.
The Regge background is simpler in this case since the only allowed Regge exchange 
is $\rho$ and we have taken
for the generic amplitude an expression of the form Eq.~(\ref{ampreg}),
with the same Regge trajectory as before, only the Regge residue being different.

Finally in the case of $Kp$ elastic scattering, for the Pomeron exchange 
the procedure is the same as for $\pi p$ and the use of an expression like Eq. (\ref{formfp}) 
leads to the introduction of {\it three} further free parameters   
$ f_{K}$, $m_{3K}$, and $ a_{K}$. However the Regge background is a little bit
more complicated, since the allowed Regge exchanges are $A_2$, $\rho$, $\omega$, as in the
$pp$ ($\bar p p$) reaction. The same Regge trajectories will be used and again
only the Regge residues are different.

To summarize, for all the cases we are considering, the elastic scattering amplitude 
reduces to a Bessel transform 
\beq{bessamp}
a(s,t) = is \int_{0}^{\infty} J_0(b\sqrt{-t}) (1 - 
e^{-\Omega_0(s,\mbf{b})}) b d{b} \ .
\eeq
We define the ratio of real to imaginary parts of the forward amplitude
\beq{ratioro}
\rho (s) = {\mbox{Re}~a(s, t=0) \over \mbox{Im}~a(s, t=0)} \ ,
\eeq
the total cross section
\beq{sigtot}
\sigma_{tot} (s) = \frac{4\pi}{s} \mbox{Im}~a(s, t=0) \ 
\eeq
and the differential cross section
\beq{dsigdt}
{d\sigma (s,t) \over dt} = \frac{\pi}{s^2}|a(s,t)|^2 \ .
\eeq

This completes the description of the scattering amplitudes and we now turn to
the phenomenological results and predictions for future experiments.

\newpage
%%%%%%%%%%%%%%%%%%%%%%%%%%%%%%%%%%%%%%%%%%%%%%%%%%%%%%%%%%%%%%%%%%%%%%%%%%%%%%%%%%%%
\section{\bf Phenomenological results and predictions}
\label{num}
\begin{figure}[ht]
\epsfxsize=12cm
\epsfysize=16cm
\centerline{\epsfbox{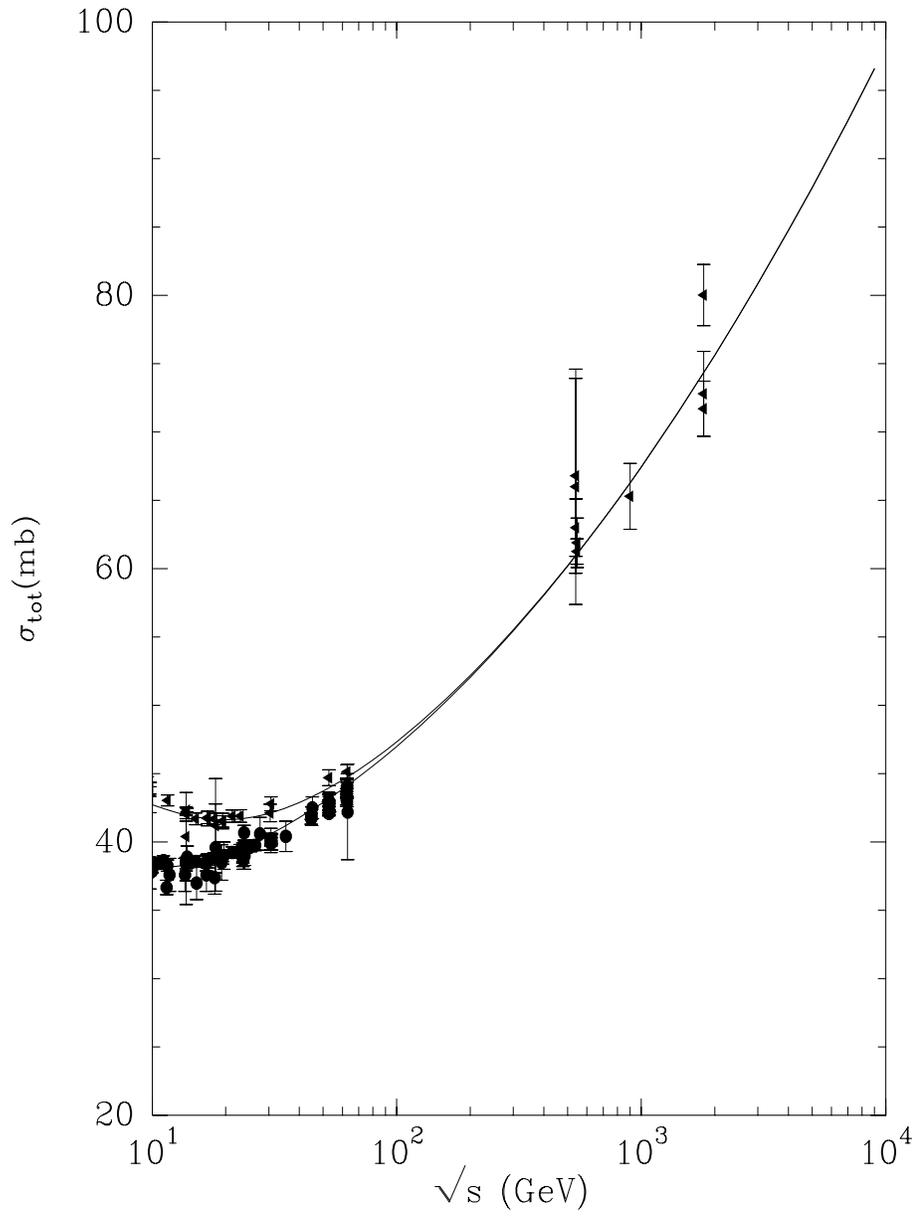}}
\caption{$\sigma_{tot}$ for $\bar p p$ and $p p$ as a function of $\sqrt{s}$.
Data compilation from Ref. \cite{PDG}.}
\label{fi:sigp}
\end{figure}

\begin{figure}[ht]
\epsfxsize=12cm
\epsfysize=16cm
\centerline{\epsfbox{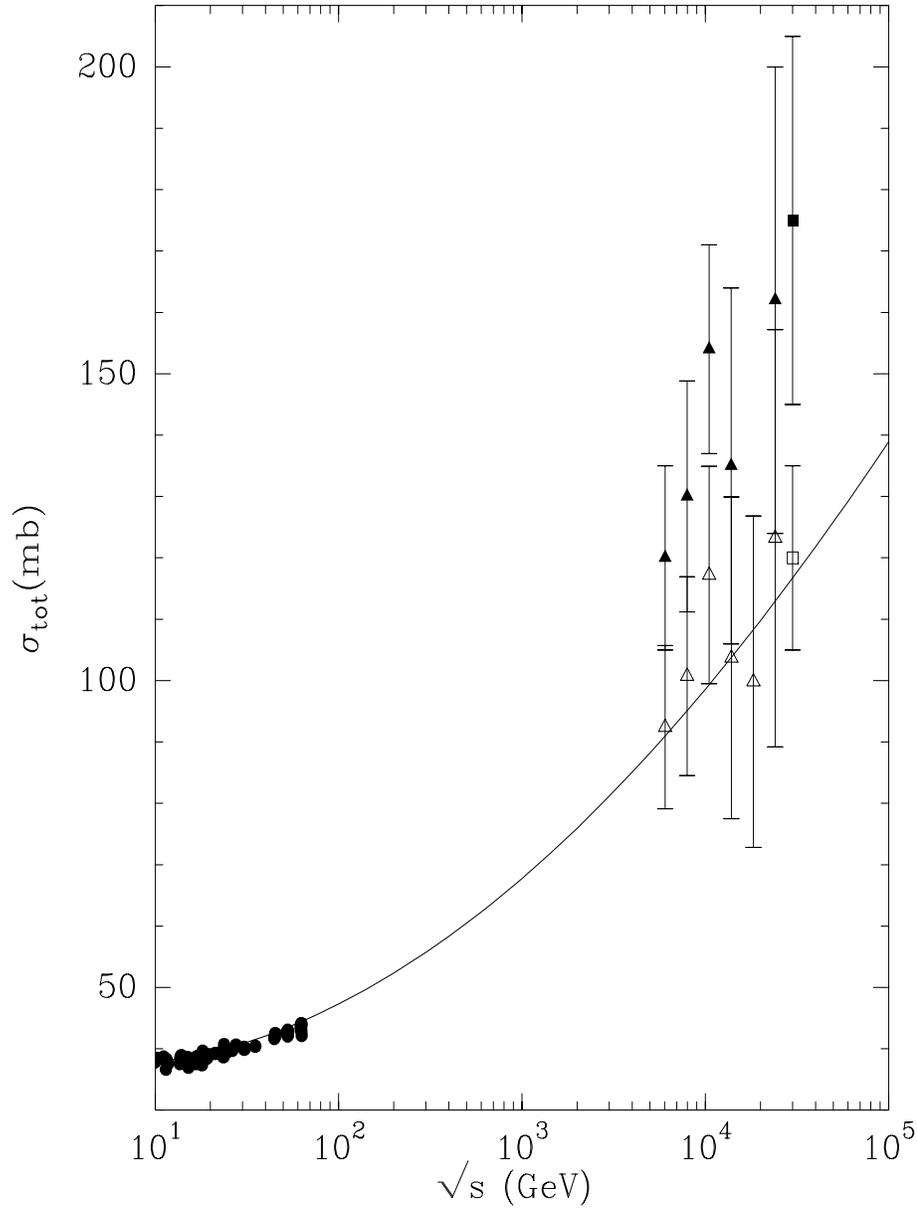}}
\caption{Comparison of $\sigma_{tot}$ for $p p$ as a function of $\sqrt{s}$
with cosmic-ray experiments.
Data from Refs. \cite{balt84,gais87,hond93,nik93}.}
\label{fi:cosmic}
\end{figure}

\begin{figure}[ht]
\epsfxsize=12cm
\epsfysize=16cm
\centerline{\epsfbox{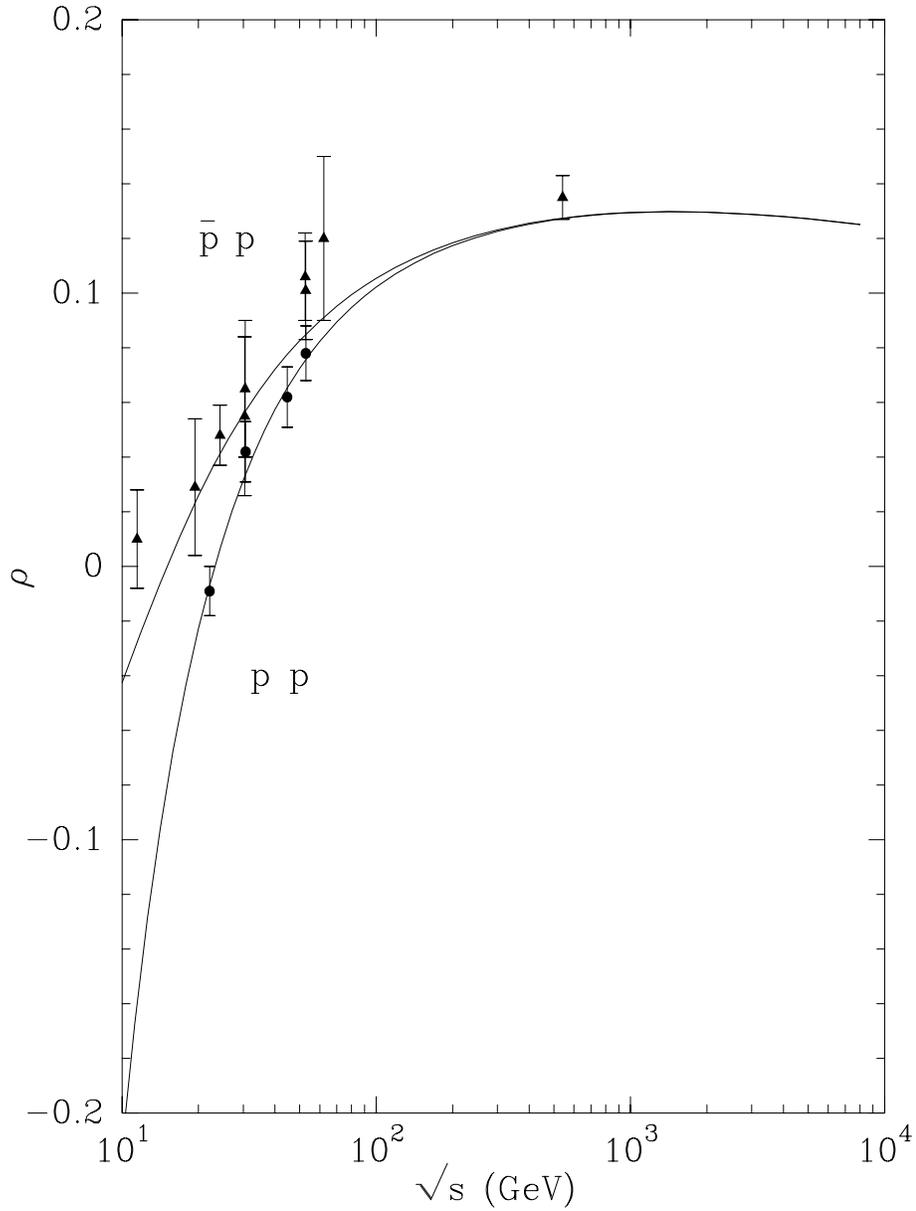}}
\caption{$\rho_{pp}$ and $\rho_{\bar p p}$ as a function of $\sqrt{s}$.
Data compilation from Ref. \cite{PDG} (full triangles 
$\bar pp$ and full circles $pp$).}
\label{fi:combrop}
\end{figure}

\begin{figure}[ht]
\epsfxsize=12cm
\epsfysize=16cm
\centerline{\epsfbox{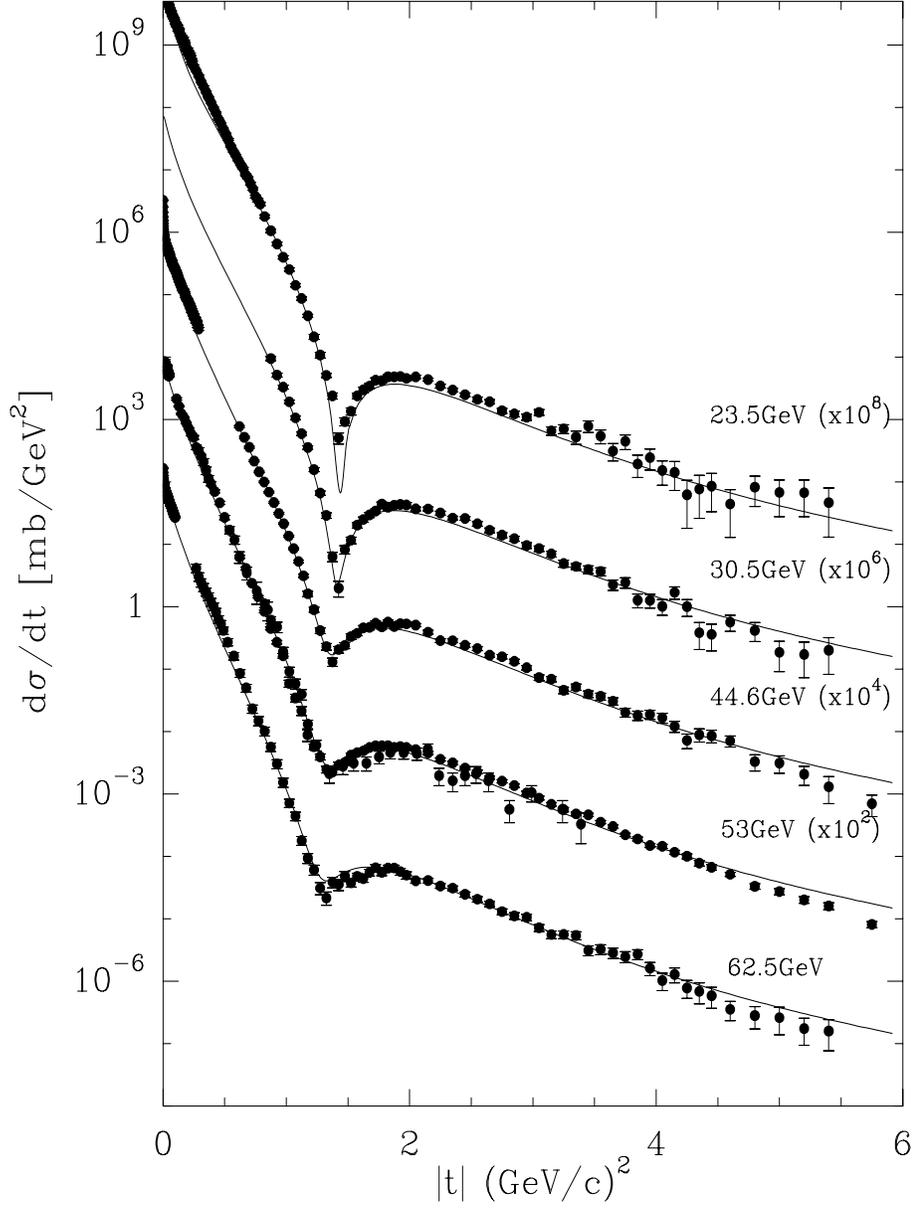}}
\caption{$d\sigma /dt$ for $p p$ as a function of $|t|$ for 
$\sqrt{s}= 23.5, 30.5, 44.8, 53, 62.5 \mbox{GeV}$. 
Experiments from Refs.
\cite{kwak,amal79b,nagy79a,break85,ambro82,break84}.}
\label{fi:dsigpp}
\end{figure}

\begin{figure}[ht]
\epsfxsize=12cm
\epsfysize=16cm
\centerline{\epsfbox{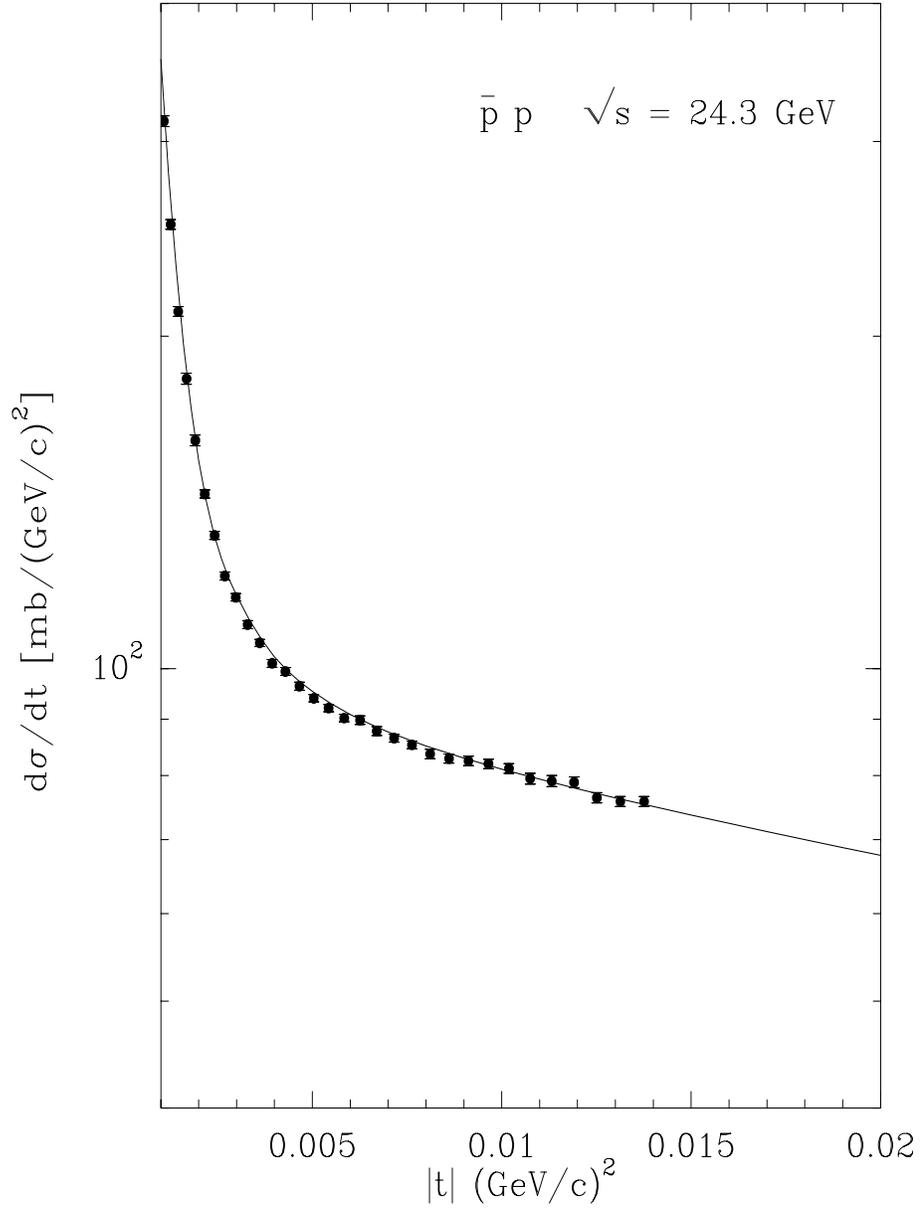}}
\caption{$d\sigma /dt$ for $\bar p p$ as a function of $|t|$ 
in the small $t$ region for 
$\sqrt{s}= 24.3 \mbox{GeV}$. Experiment UA6 \cite{bree89}.}
\label{dsigpbp24}
\end{figure}

\begin{figure}[ht]
\epsfxsize=12cm
\epsfysize=16cm
\centerline{\epsfbox{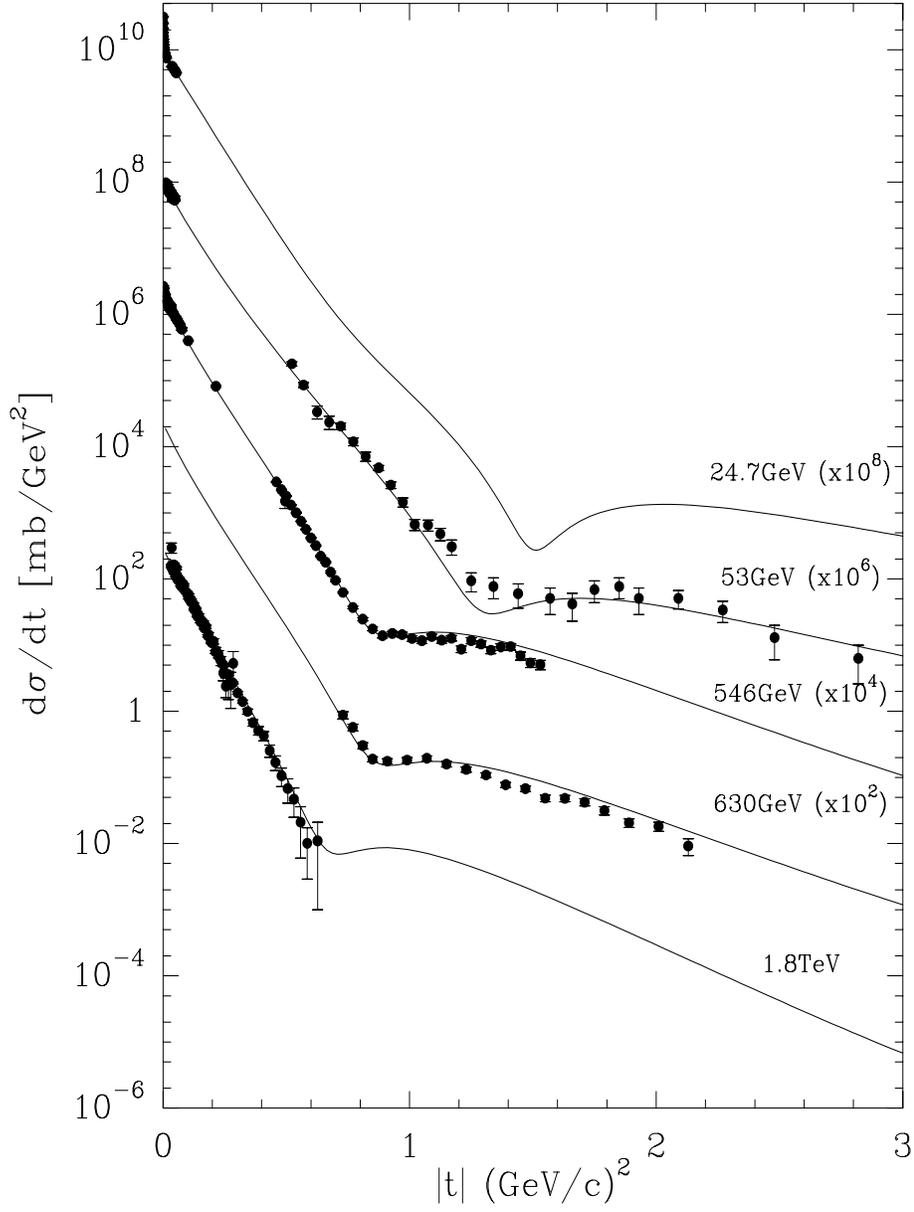}}
\caption{$d\sigma /dt$ for $\bar p p$ as a function of $|t|$ for 
$\sqrt{s}= 24.7, 53, 546, 630, 1800 \mbox{GeV}$. 
Experiments from Refs.
\cite{break85,bree89,ambro82,aug93,abe93,boz85,bern86}.}
\label{fi:dsigpbp}
\end{figure}

\begin{figure}[ht]
\epsfxsize=12cm
\centerline{\epsfbox{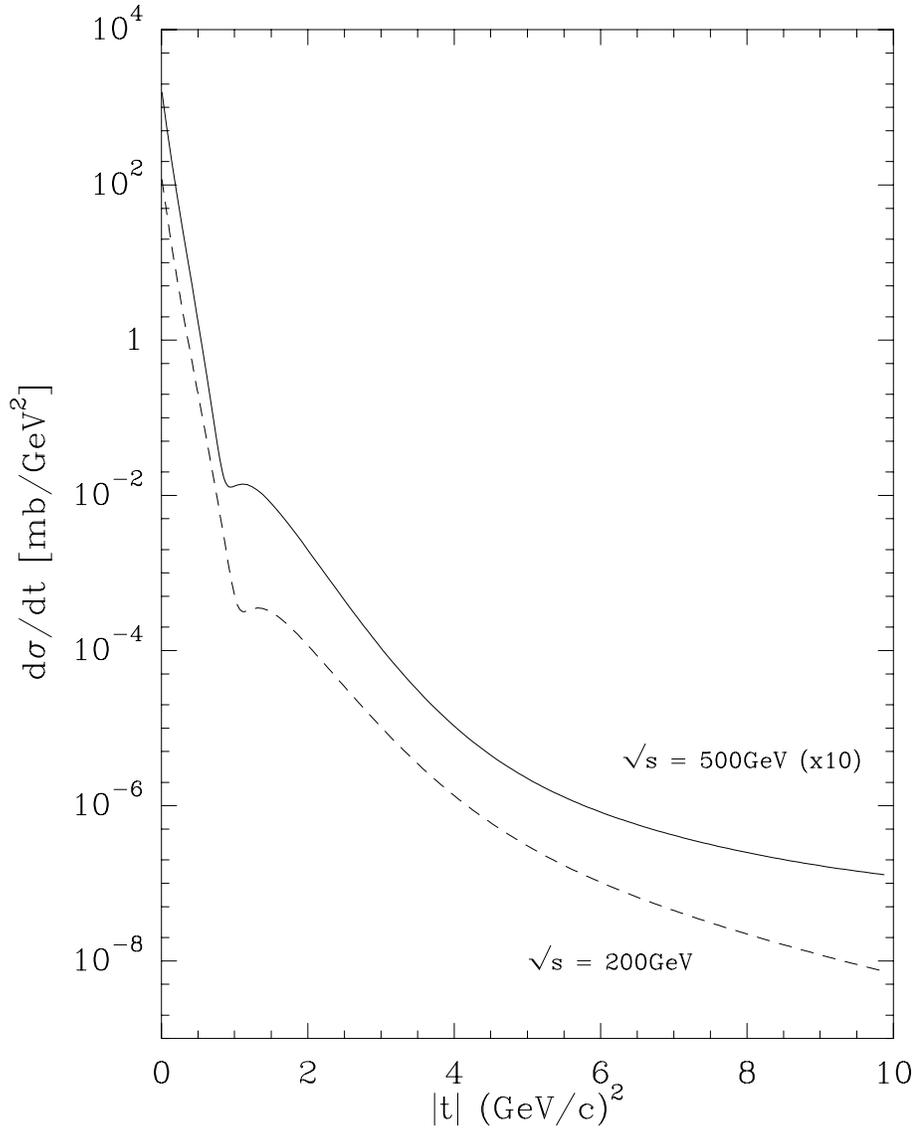}}
\caption{Predicted $d\sigma /dt$ for $pp$ as a function of 
$|t|$ for the BNL-RHIC energy domain.}
\label{fi:rhicpp}
\end{figure}

\begin{figure}[ht]
\epsfxsize=12cm
\centerline{\epsfbox{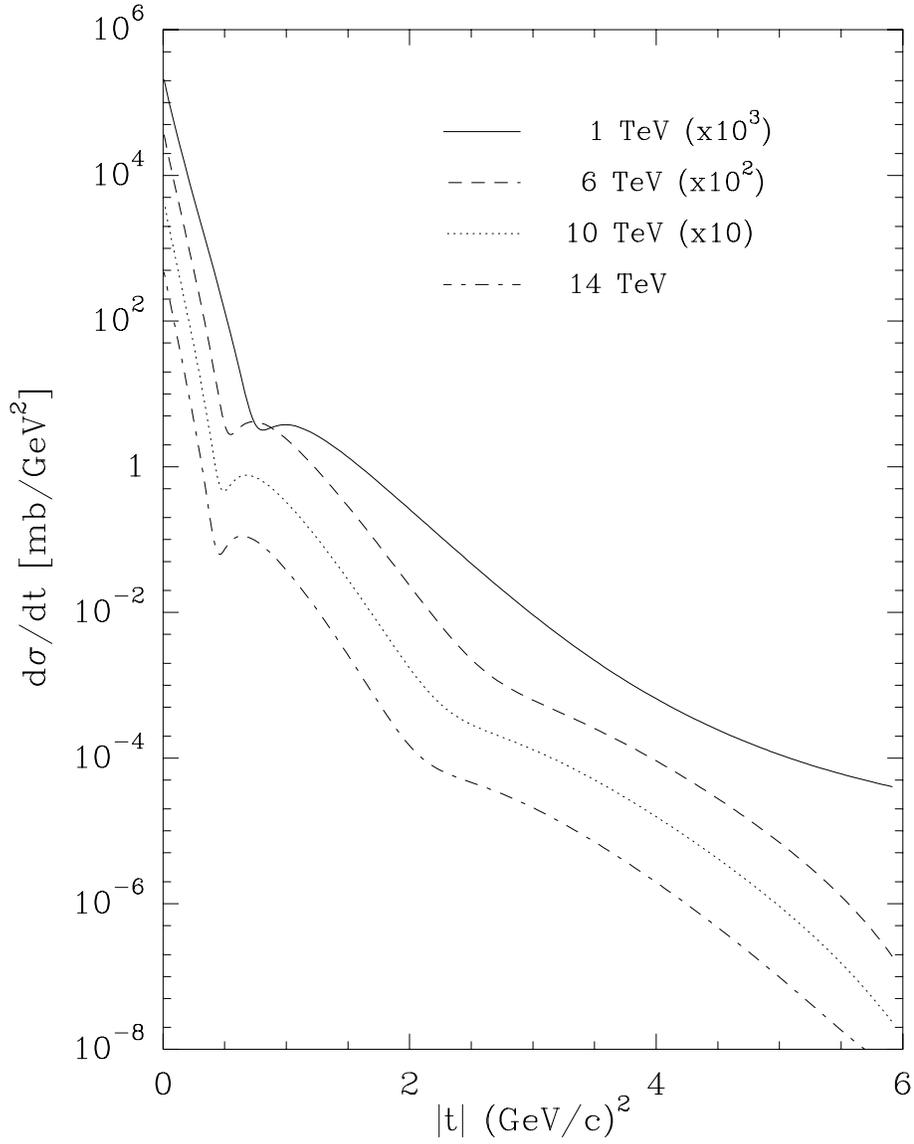}}
\caption{Predicted $d\sigma /dt$ for $p p$ as a function of 
$|t|$ for different $\sqrt{s}$ in the TeV energy domain.}
\label{fi:dsigpptev}
\end{figure}

We have made a global fit for all the elastic reactions under study
and for all of them we have introduced a lower cutoff 
at $p_{lab} = 100\mbox{Gev/c}$.
When making the fit we have analyzed high energy experimental $pp$ ($\bar p p$) data including
a set of 369 points coming from ISR, SPS and Tevatron experiments on 
$\rho(s)$, 
$\sigma_{tot} (s)$ and  $d\sigma (s,t) / dt$. The new $pp$ ($\bar p p$) Pomeron parameters are
listed in Table \ref{table1} together with those obtained in our previous 
analysis.
We notice that these four parameters have changed only slightly, within a few percents.
The $C$ values of the Regge residues are given in Table~\ref{table2a}.
The parameters $b$ are zero except for the $\rho$ contribution for which we found
$b_{\rho} = 8.54$. 
Let us notice that the Regge exchanges have been eikonalized so
the intercept and the slope of the trajectories are 
effective parameters whose values are given in Table~\ref{table2b}. However we 
notice that the three intercepts are close to 1/3 and the $A_2$ and $\rho$ slopes
have the standard value $\alpha ^{'}$= 1GeV$^{-2}$, whereas the $\omega$ slope is
slightly smaller.
A detailed $\chi^2$ analysis is presented in Table~\ref{table3}, where
the present solution which is a fit to the data is compared to the 1984 
solution,  
which is a prediction made with the same set of data. We notice a substantial
improvement in the $\chi^2$, since we have now a $\chi^2/\mbox{pt} = 4.45$.
A comparison of our model with experimental data is 
given in the Figs. \ref{fi:sigp}-\ref{fi:dsigpbp}. In 
Fig. \ref{fi:rhicpp} and Fig. \ref{fi:dsigpptev},
we display some predictions for future measurements at BNL-RHIC and CERN-LHC,
respectively.

\begin{figure}[htbp]
\epsfxsize=12cm
\epsfysize=16cm
\centerline{\epsfbox{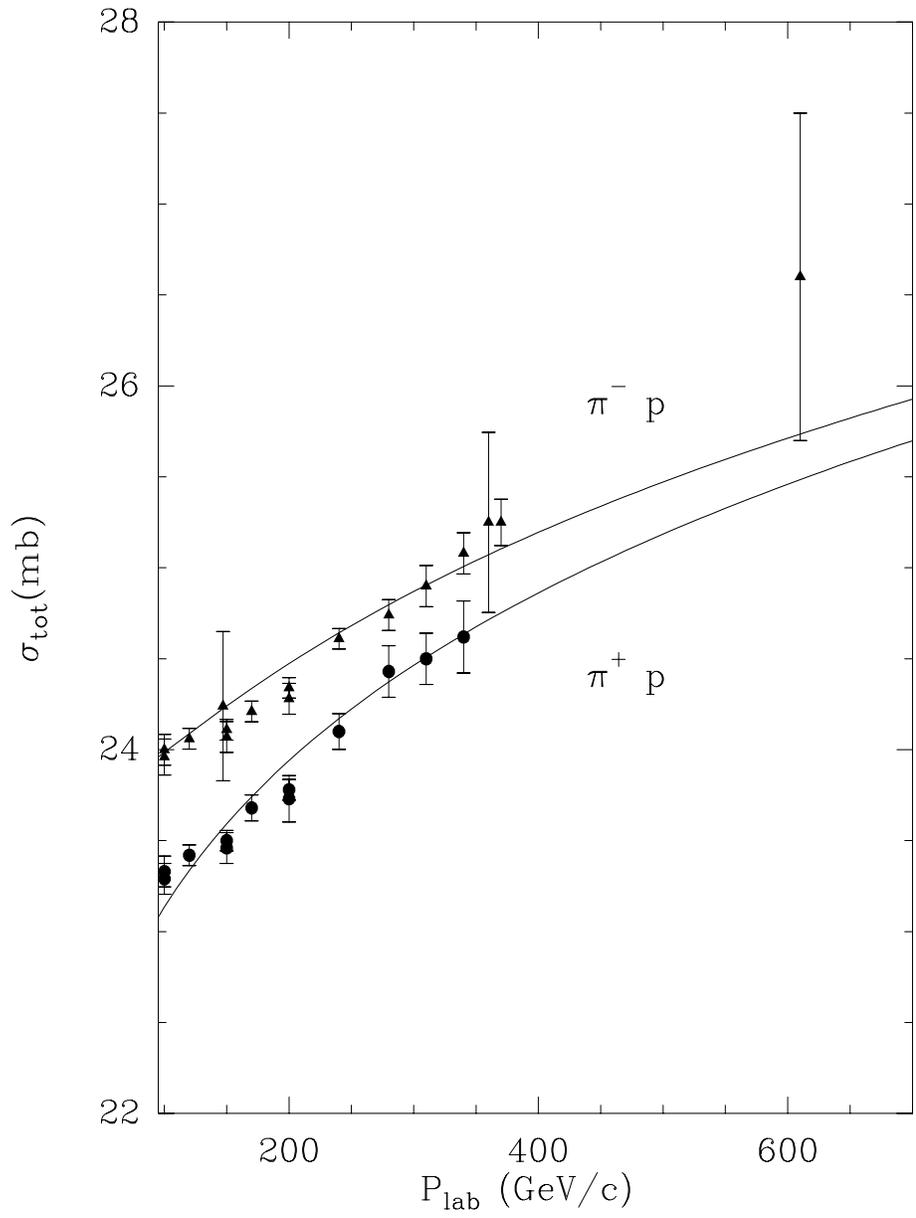}}
\caption{$\sigma_{tot}$ for $\pi^{\pm} p$ as a function of $p_{lab}(\mbox{GeV/c})$.
Data compilation from Ref. \cite{PDG}.}
\label{fi:sigpi}
\end{figure}

\begin{figure}[htbp]
\epsfxsize=12cm
\epsfysize=16cm
\centerline{\epsfbox{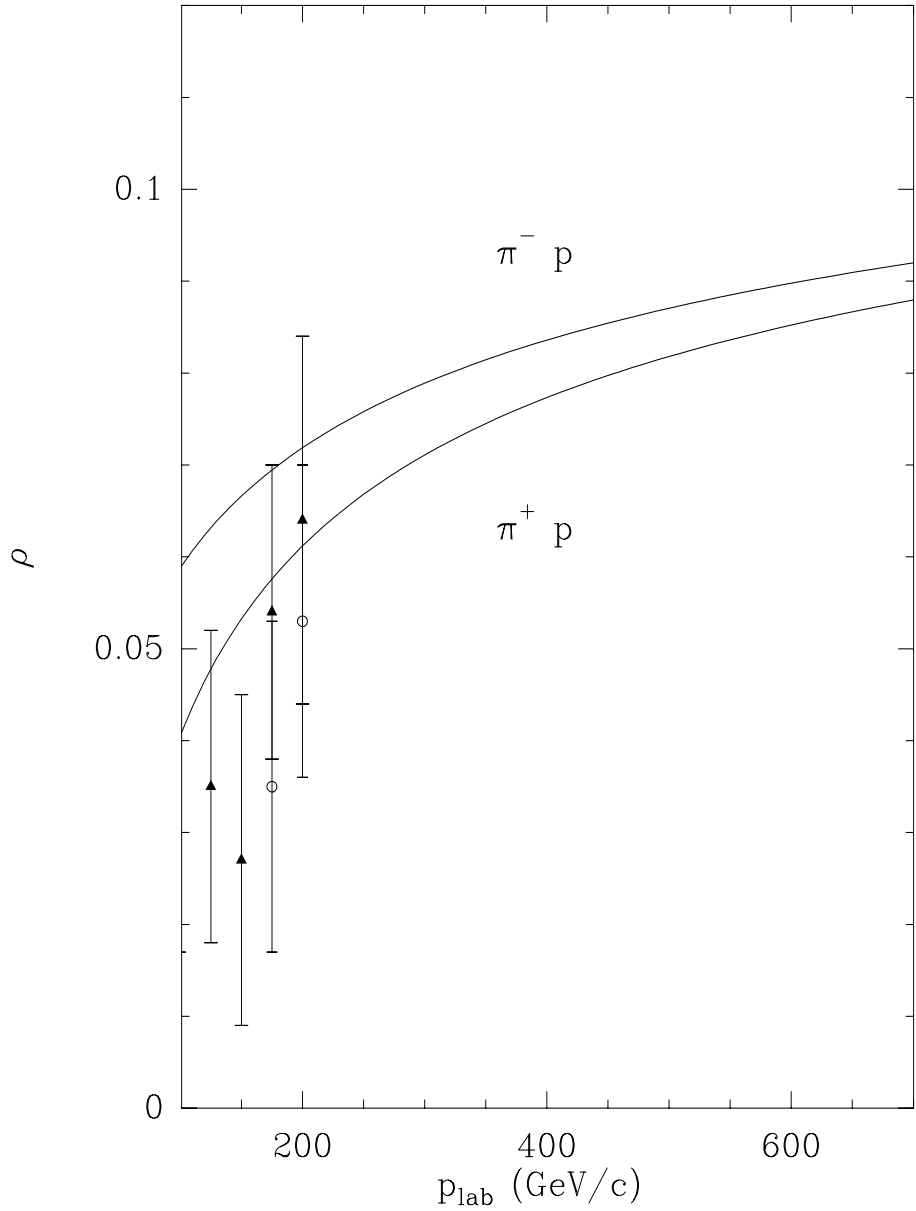}}
\caption{$\rho$ for $\pi^{\pm} p$ as a function of $p_{lab}(\mbox{GeV/c})$
($\pi^-$ triangles, $\pi^+$ open circles).
Data compilation from Ref. \cite{PDG}.}
\label{fi:combropi}
\end{figure}

\begin{figure}[htbp]
\epsfxsize=12cm
\epsfysize=16cm
\centerline{\epsfbox{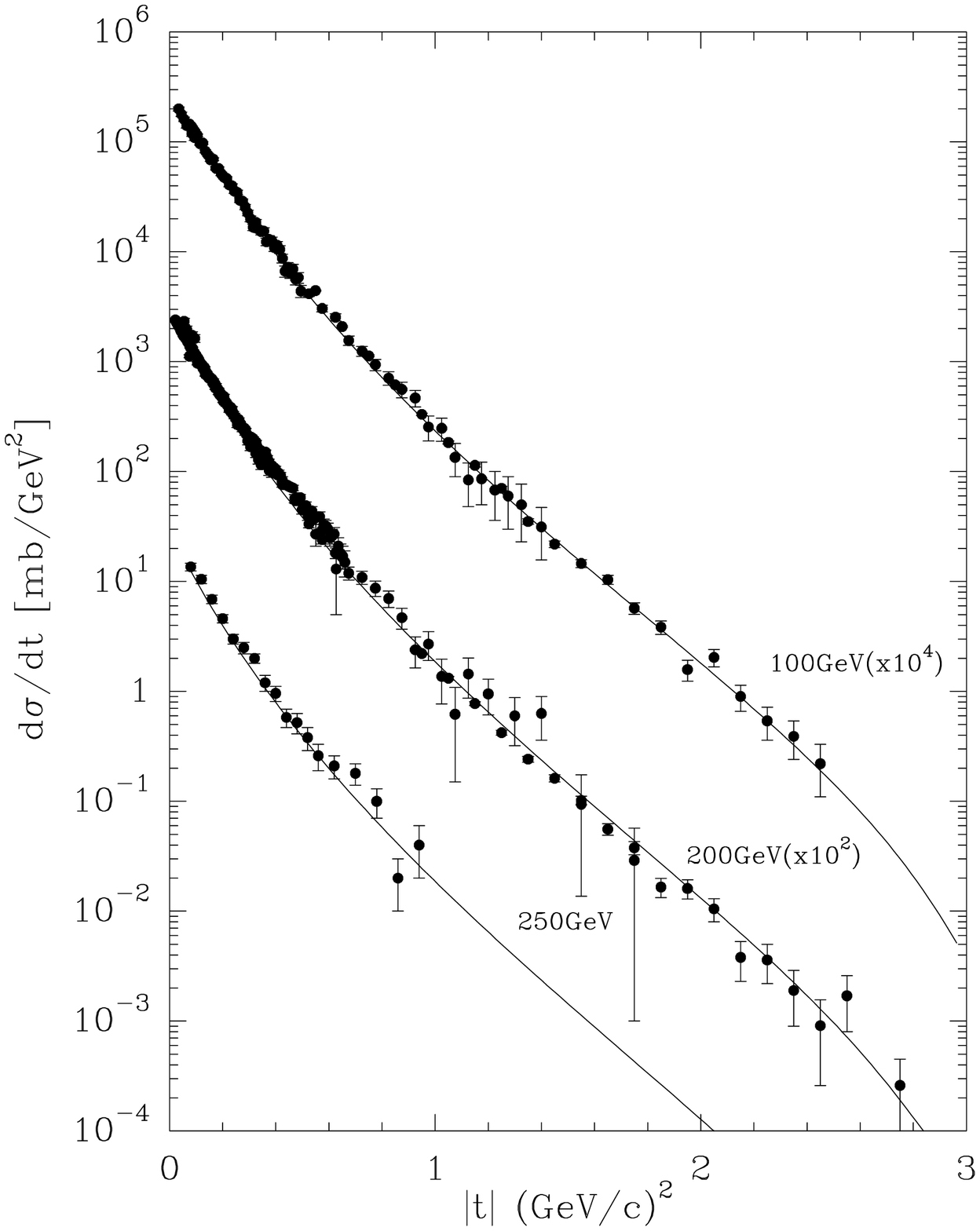}}
\caption{$d\sigma /dt$ for $\pi^+ p$ as a function of $|t|$ for 
$p_{lab} = 100, 200, 250 \mbox{GeV/c}$. 
Experiments from Refs.  \cite{aker76,rubi84,cool81,schiz79,adam87}.}
\label{fi:dsigpi+p}
\end{figure}

\begin{figure}[htbp]
\epsfxsize=12cm
\epsfysize=16cm
\centerline{\epsfbox{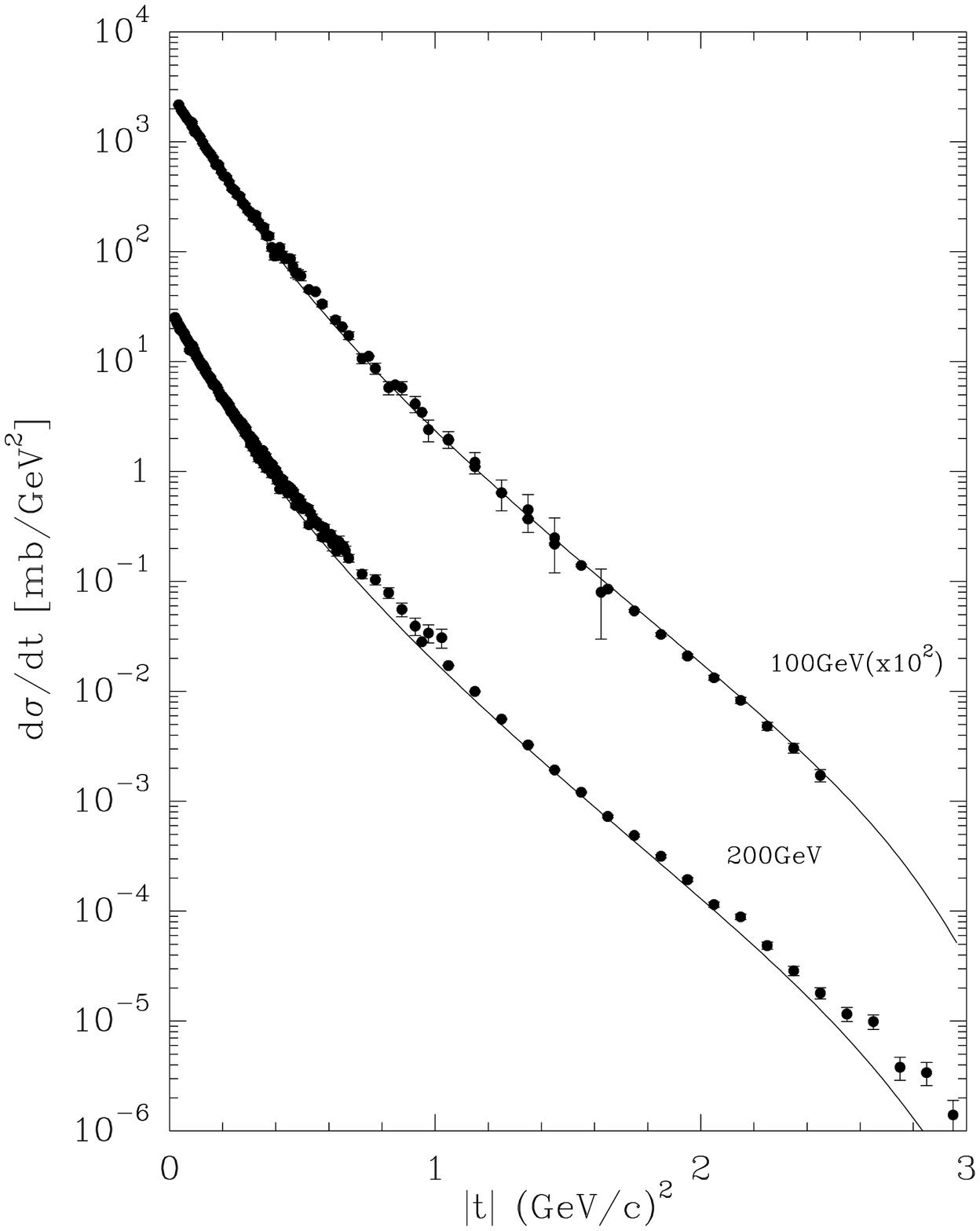}}
\caption{$d\sigma /dt$ for $\pi^- p$ as a function of $|t|$ for 
$p_{lab} = 100, 200 \mbox{GeV/c}$. Experiments from Refs.
\cite{aker76,rubi84,cool81,schiz79}.}
\label{fi:dsigpi-p}
\end{figure}

\begin{table}[ht]
\begin{center}
\begin{tabular}{|c|c|c|}
\hline
&&\\
year & present & 1984\\
&&\\
%\hline\raisebox{0pt}[12pt][6pt]{c}    &0.167560    & 0.167560  \\[4pt]
%\hline\raisebox{0pt}[12pt][6pt]{c'}   &0.747970    & 0.747970  \\[4pt]
\hline\raisebox{0pt}[12pt][6pt]{$m_1$}&0.577225    & 0.586  \\[4pt]
\hline\raisebox{0pt}[12pt][6pt]{$m_2$}&1.719896    & 1.704  \\[4pt]
\hline\raisebox{0pt}[12pt][6pt]{$f$}  &6.970913   & 7.115  \\[4pt]
\hline\raisebox{0pt}[12pt][6pt]{a}    &1.858442    & 1.953  \\[4pt]
\hline
\end{tabular}
\caption {Pomeron parameters for $pp$ ($\bar pp$). Comparison of the present 
and 1984 solutions.}
\label{table1}
\end{center}
\end{table}
%%%%%%%%%%%%%%
\newpage
\clearpage
\begin{table}[ht]
\begin{center}
\begin{tabular}{|c|c|}
\hline
exchange & $C$   \\
\hline\raisebox{0pt}[12pt][6pt]{$A_2$}    & -24.2686  \\[4pt]
\hline\raisebox{0pt}[12pt][6pt]{$\omega$}& -167.3293\\[4pt]
\hline\raisebox{0pt}[12pt][6pt]{$\rho$}& 124.91969  \\[4pt]
\hline
\end{tabular}
\caption{$pp$ ($\bar pp$) Regge parameters.}
\label{table2a}
\vspace{3pt}
\begin{tabular}{|c|c|c|c|}
\hline
trajectory & $A_2$ & $\omega$ & $\rho$ \\
\hline\raisebox{0pt}[12pt][6pt] 
&$0.3566 + t$ & $0.3229 + 0.7954 t$ & $0.3202 + t$ \\[4pt]
\hline
\end{tabular}
\caption{Effective Regge trajectories.}
\label{table2b}
\vspace{3pt}
\begin{tabular}{|c|c|c|c|}
\hline
&&&\\
process & $\chi^2$ present & $\chi^2$ 1984 & nb points\\
&&&\\
\hline\raisebox{0pt}[12pt][6pt]
{$d\sigma (pp)/dt$ $p_{lab}=280\mbox{GeV/c}$}    & 351    & 538 & 58  \\[4pt]
\hline\raisebox{0pt}[12pt][6pt]
{$d\sigma (pp)/dt$ $p_{lab}=496\mbox{GeV/c}$}    & 194    & 527 & 54  \\[4pt]
\hline\raisebox{0pt}[12pt][6pt]
{$d\sigma (pp)/dt$ $p_{lab}=1486\mbox{GeV/c}$}    & 246    & 190 & 16  \\[4pt]
\hline\raisebox{0pt}[12pt][6pt]
{$d\sigma (pp)/dt$ $p_{lab}=2060\mbox{GeV/c}$}    & 40    & 42 & 10  \\[4pt]
\hline\raisebox{0pt}[12pt][6pt]
{$d\sigma (pp)/dt$ $p_{lab}=2080\mbox{GeV/c}$}    & 97    & 135 & 51  \\[4pt]
\hline\raisebox{0pt}[12pt][6pt]
{$d\sigma (\bar pp)/dt$ $p_{lab}=315\mbox{GeV/c}$} & 51 & 516 & 41  \\[4pt]
\hline\raisebox{0pt}[12pt][6pt]
{$d\sigma (\bar pp)/dt$ $\sqrt{s}=546\mbox{GeV}$} & 474 & 1190 & 84  \\[4pt]
\hline\raisebox{0pt}[12pt][6pt]
{$d\sigma (\bar pp)/dt$ $\sqrt{s}=630\mbox{GeV}$} & 141 & 628 & 19  \\[4pt]
\hline\raisebox{0pt}[12pt][6pt]
{$\sigma_{tot} (pp)$ } & 16 & 144 & 13  \\[4pt]
\hline\raisebox{0pt}[12pt][6pt]
{$\sigma_{tot} (\bar pp)$ } & 21 & 44 & 10  \\[4pt]
\hline\raisebox{0pt}[12pt][6pt]
{$\rho (pp)$ } & 2 & 9 & 7  \\[4pt]
\hline\raisebox{0pt}[12pt][6pt]
{$\rho (\bar pp)$ } & 8 & 11 & 6  \\[4pt]
\hline\raisebox{0pt}[12pt][6pt]
{Total } & 1641 & 3974 & 369  \\[4pt]
\hline
\end{tabular}
\caption {Detailed $\chi^2$ comparison between the present and 1984 solutions
 for $pp$
and $\bar pp$}
\label{table3}
\end{center}
\end{table}
\clearpage
For $\pi^{\pm} p$ elastic scattering, from 113 data points we have obtained the following
Pomeron parameters values :
$$ m_{3\pi} = 0.7665,~~~~~ f_{\pi} = 4.2414,~~~~~~~ a_{\pi} = 2.3272.$$
Due to the lack of very high energy data for these reactions,
the Regge contributions play a non negligible role in the fit  
we made with a cut at $p_{lab} = 100\mbox{GeV/c}$.
The only allowed Regge exchange is $\rho$ and we  
find $C_{\rho}= 4.1624$ and  $b_{\rho} = 4.2704$.
A detailed $\chi^2$ analysis of the data is given in Table~\ref{table5}
and we have a $\chi^2/\mbox{pt} = 4$.
A comparison of our model with experimental data is 
given in Figs. \ref{fi:sigpi}-\ref{fi:dsigpi-p} and in 
Fig. \ref{fi:dsigpi-ptev}
we give some predictions for future measurements at CERN-LHC. 

%%%%%%%%%%%%%%%%%%%%%%%%%%%%%%%%%%%%%%%%%%%%%%%%%%%%%
\begin{table}[htb]
\begin{center}
\begin{tabular}{|c|c|c|}
\hline
&&\\
process & $\chi^2$  & nb points\\
&&\\
\hline\raisebox{0pt}[12pt][6pt]
{$d\sigma (\pi^- p)/dt$ $p_{lab}=100\mbox{GeV/c}$}    & 112    & 32  
\\[4pt]
\hline\raisebox{0pt}[12pt][6pt]
{$d\sigma (\pi^- p)/dt$ $p_{lab}=200\mbox{GeV/c}$}    & 139    & 14  
\\[4pt]
\hline\raisebox{0pt}[12pt][6pt]
{$d\sigma (\pi^+ p)/dt$ $p_{lab}=100\mbox{GeV/c}$}    & 50    & 32  
\\[4pt]
\hline\raisebox{0pt}[12pt][6pt]
{$d\sigma (\pi^+ p)/dt$ $p_{lab}=200\mbox{GeV/c}$}    & 58    & 14  
\\[4pt]
\hline\raisebox{0pt}[12pt][6pt]
{$\sigma_{tot} (\pi^- p)$ } & 64 & 8  \\[4pt]
\hline\raisebox{0pt}[12pt][6pt]
{$\sigma_{tot} (\pi^+ p)$ } & 16 & 6  \\[4pt]
\hline\raisebox{0pt}[12pt][6pt]
{$\rho (\pi^- p)$ } & 8 & 4  \\[4pt]
\hline\raisebox{0pt}[12pt][6pt]
{$\rho (\pi^+ p)$ } & 6 & 3  \\[4pt]
\hline\raisebox{0pt}[12pt][6pt]
{Total } & 453 & 113  \\[4pt]
\hline
\end{tabular}
\caption {Detailed $\chi^2$ for $\pi p$ fitted processes}
\label{table5}
\end{center}
\end{table}

\begin{figure}[htbp]
\epsfxsize=12cm
\epsfysize=16cm
\centerline{\epsfbox{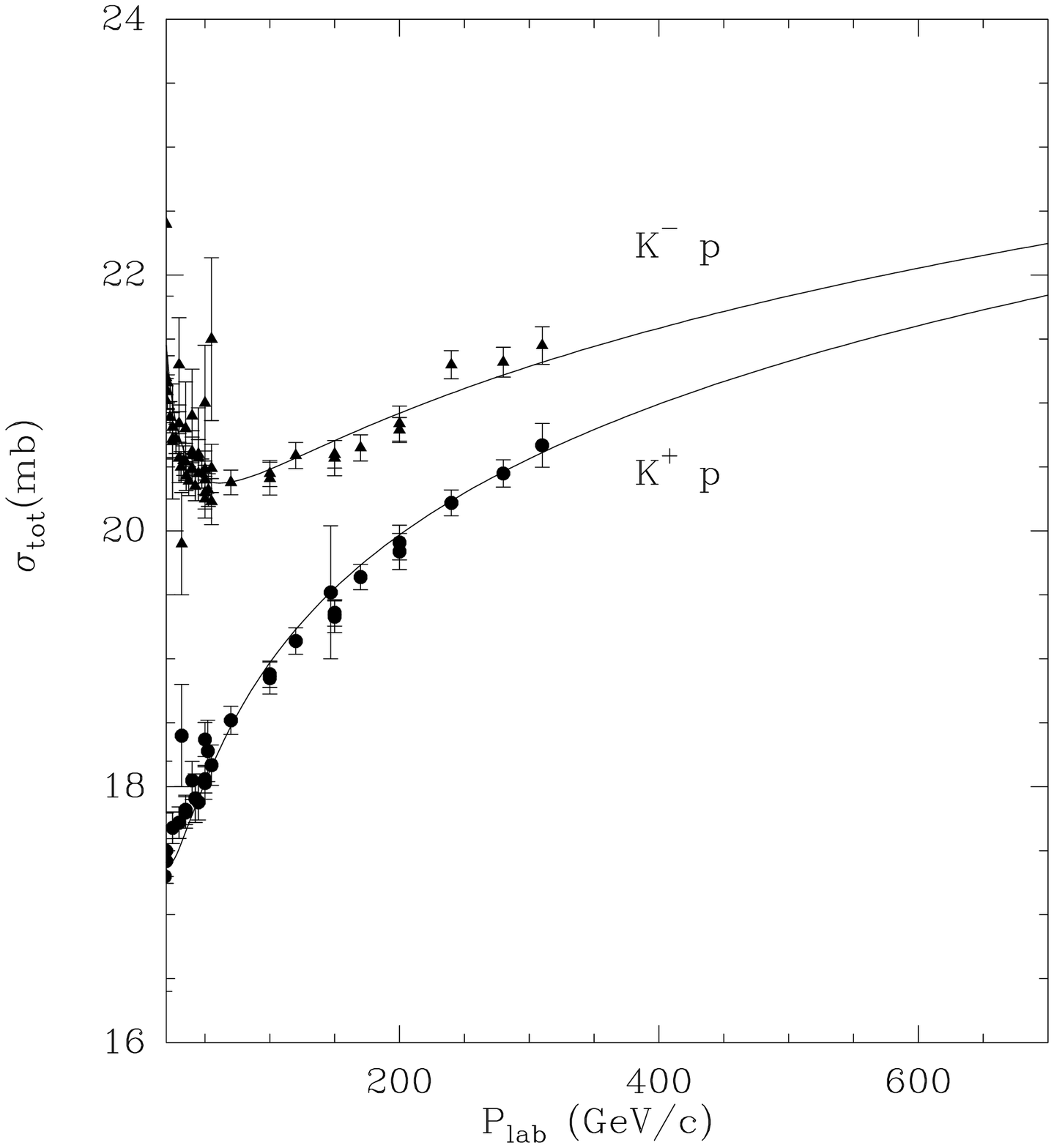}}
\caption{$\sigma_{tot}$ for $K^{\pm} p$ as a function of $p_{lab}(\mbox{GeV/c})$,
compilation of data from Ref. \cite{PDG}.}
\label{fi:sigk}
\end{figure}

\begin{figure}[htbp]
\epsfxsize=12cm
\epsfysize=16cm
\centerline{\epsfbox{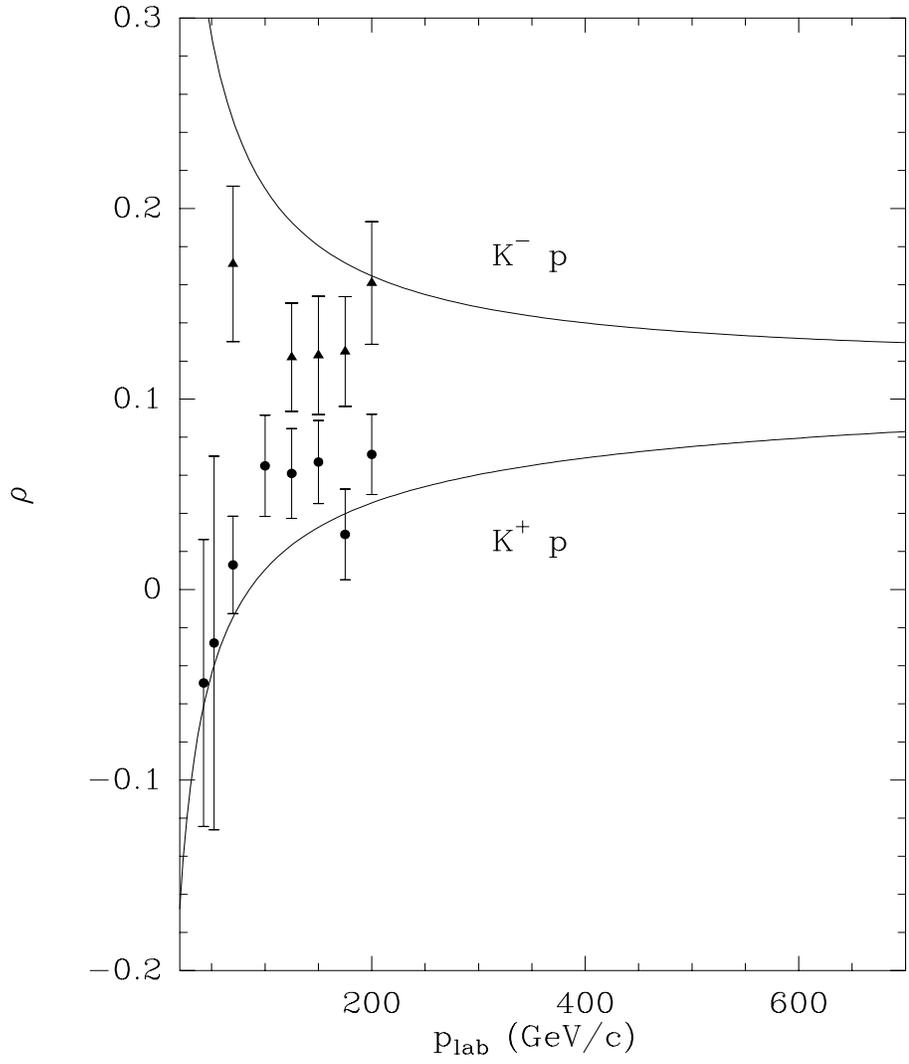}}
\caption{$\rho$ for $K^{\pm} p$ as a function of $p_{lab}(\mbox{GeV/c})$,
($K^-$ triangles, $K^+$ circles).
Compilation of data from Ref. \cite{PDG}.}
\label{fi:combrok}
\end{figure}

\begin{figure}[ht]
\epsfxsize=12cm
\epsfysize=16cm
\centerline{\epsfbox{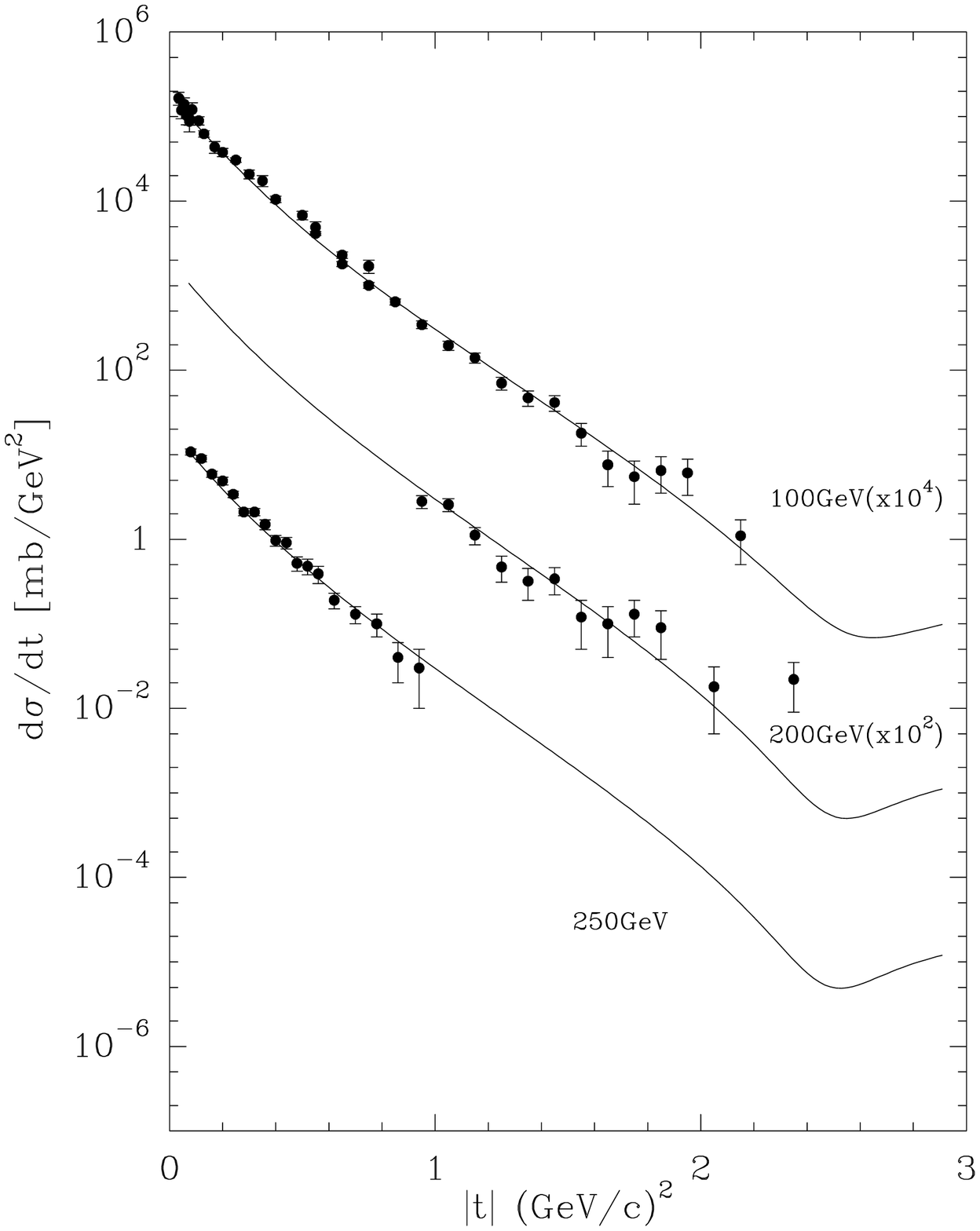}}
\caption{$d\sigma /dt$ for $K^{+} p$ as a function of $|t|$ for 
$p_{lab}= 100, 200, 250 \mbox{GeV/c}$. 
Experiments from Refs. \cite{rubi84,cool81,adam87,ayres76}.}
\label{fi:dsigk+p}
\end{figure}

\begin{figure}[ht]
\epsfxsize=12cm
\epsfysize=16cm
\centerline{\epsfbox{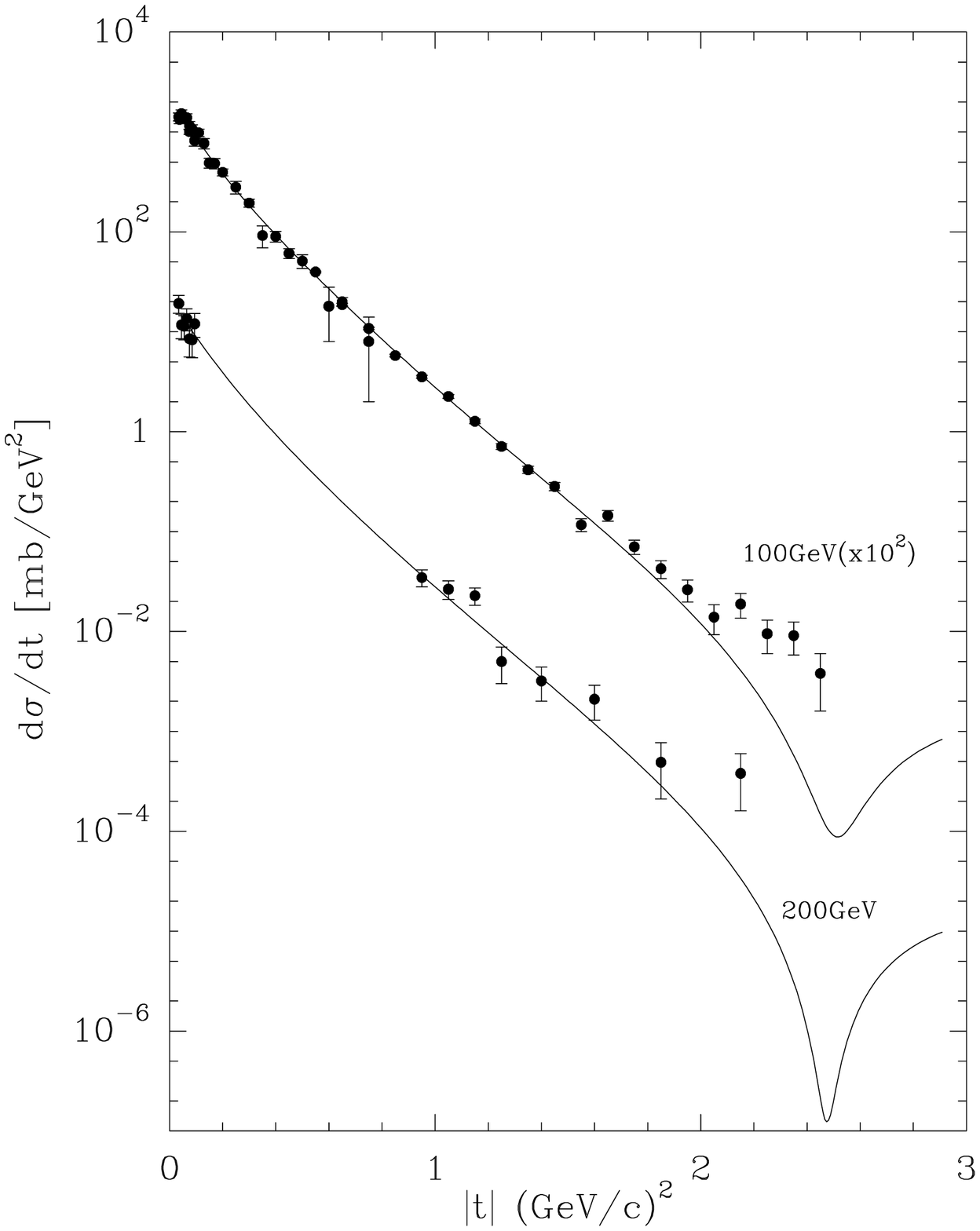}}
\caption{$d\sigma /dt$ for $K^{-} p$ as a function of $|t|$ for 
$p_{lab}= 100, 200 \mbox{GeV/c}$. 
Experiments from Refs. \cite{rubi84,cool81,ayres76}.}
\label{fi:dsigk-p}
\end{figure}

Finally for $K^{\pm}p$ elastic scattering where we used 212 data points, the procedure is the same 
as in the $\pi^{\pm}p$ case and the Pomeron parameters obtained from the fit are
$$ m_{3K} = 1.1391,~~~~~ f_{K} = 3.6729,~~~~~~~ a_{K} = 1.9913.$$
For these reactions the allowed exchanges are the $A_2$, $\rho$ and
$\omega$. We have kept the same trajectories 
as in the $p p$ case, with $b_{A2} = b_{\omega} = 8.5409$, and $b_{\rho} = 4.2704$ and
the values of the corresponding C parameters are listed in
Table~\ref{table6}.
A detailed $\chi^2$ analysis of the data is given in Table~\ref{table7}
and we have a $\chi^2/\mbox{pt} = 1.5$.
A comparison of our model with experimental data is 
given in Figs. \ref{fi:sigk}-\ref{fi:dsigk-p} and in 
Fig. \ref{fi:dsigk-ptev}
we give some predictions for future measurements at CERN-LHC.
It might be possible to expect future measurements at BNL-RHIC with 
extracted $\pi^{\pm}$, $K^{\pm}$ beams, to improve the available data. 
\newpage
%%%%%%%%%%%%%%
\begin{table}[hb]
\begin{center}
\begin{tabular}{|c|c|}
\hline
exchange & $C$  \\
\hline\raisebox{0pt}[12pt][6pt]{$A_2$}    & 6.3221 \\[4pt]
\hline\raisebox{0pt}[12pt][6pt]{$\rho$}& -39.4989  \\[4pt]
\hline\raisebox{0pt}[12pt][6pt]{$\omega$}& 46.5386 \\[4pt]
\hline
\end{tabular}
\caption {$K p$ Regge parameters.}
\label{table6}
\vspace{4pt}
\begin{tabular}{|c|c|c|}
\hline
&&\\
process & $\chi^2$  & nb points\\
&&\\
\hline\raisebox{0pt}[12pt][6pt]
{$d\sigma (K^- p)/dt$ $p_{lab}=100\mbox{GeV/c}$}    & 120    & 43  
\\[4pt]
\hline\raisebox{0pt}[12pt][6pt]
{$d\sigma (K^- p)/dt$ $p_{lab}=140\mbox{GeV/c}$}    & 16    & 13  
\\[4pt]
\hline\raisebox{0pt}[12pt][6pt]
{$d\sigma (K^- p)/dt$ $p_{lab}=175\mbox{GeV/c}$}    & 23    & 15  
\\[4pt]
\hline\raisebox{0pt}[12pt][6pt]
{$d\sigma (K^- p)/dt$ $p_{lab}=200\mbox{GeV/c}$}    & 15    & 15  
\\[4pt]
\hline\raisebox{0pt}[12pt][6pt]
{$d\sigma (K^+ p)/dt$ $p_{lab}=100\mbox{GeV/c}$}    & 56    & 36  
\\[4pt]
\hline\raisebox{0pt}[12pt][6pt]
{$d\sigma (K^+ p)/dt$ $p_{lab}=140\mbox{GeV/c}$}    & 11    & 16  
\\[4pt]
\hline\raisebox{0pt}[12pt][6pt]
{$d\sigma (K^+ p)/dt$ $p_{lab}=175\mbox{GeV/c}$}    & 16    & 17  
\\[4pt] 
\hline\raisebox{0pt}[12pt][6pt]
{$d\sigma (K^+ p)/dt$ $p_{lab}=200\mbox{GeV/c}$}    & 17    & 12  
\\[4pt]
\hline\raisebox{0pt}[12pt][6pt]
{$\sigma_{tot} (K^- p)$ } & 13 & 17  \\[4pt]
\hline\raisebox{0pt}[12pt][6pt]
{$\sigma_{tot} (K^+ p)$ } & 8 & 15  \\[4pt]
\hline\raisebox{0pt}[12pt][6pt]
{$\rho (K^- p)$ } & 16 & 5  \\[4pt]
\hline\raisebox{0pt}[12pt][6pt]
{$\rho (K^+ p)$ } & 7 & 8  \\[4pt]
\hline\raisebox{0pt}[12pt][6pt]
{Total } & 318 & 212  \\[4pt]
\hline
\end{tabular}
\caption {Detailed $\chi^2$ for $K p$ fitted processes}
\label{table7}
\end{center}
\end{table}
\clearpage
\section{Concluding remarks}In one of the earliest 
phenomenological analyses \cite{r7} that incorporated the theoretical predictions of 
increasing cross sections \cite{r1}, results were obtained not only for $pp$ and $p\bar 
p$ total cross sections but also for those of $\pi^{+}p$, $\pi^{-}p$, $K^{+}p$ 
and $K^{-}p$.  Since then, efforts in this direction have been concentrated on 
studying the cross sections for $pp$ and $p\bar p$ scattering \cite{r8,r9,r10,r11}.

 There are perhaps two reasons why there has been a lack of progress on the cases of 
$\pi^{\pm}p$ and $K^{\pm}p$ as compared with those of $pp$ and $\bar pp$.  First, 
since the available experimental data for $\pi^{\pm}p$ and $K^{\pm}p$ scattering 
are at lower energies, the Regge backgrounds play more prominent roles, and this 
complicates the phenomenological analysis. Secondly, there has been relatively 
little reason for analyzing such cases because, while the predictions for the 
$pp$ and $\bar p p$ cases are most relevant with new experimental data expected 
at higher and higher energies every few years, up until now the availability 
of high-energy data for $\pi^{\pm}p$ and $K^{\pm}p$ elastic scattering has 
been very limited.

\begin{figure}[ht]
\epsfxsize=12cm
\epsfysize=16cm
\centerline{\epsfbox{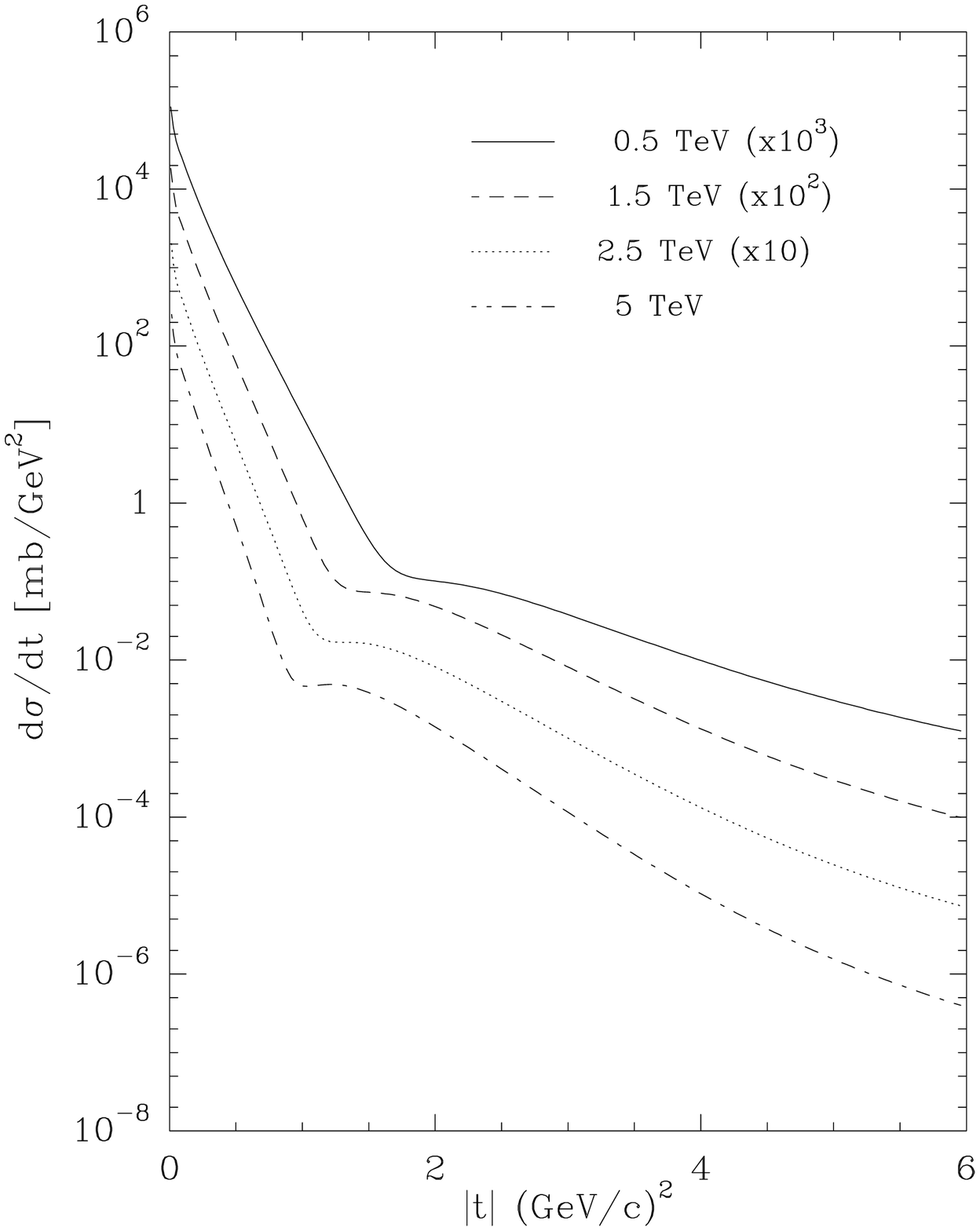}}
\caption{$d\sigma /dt$ for $\pi^- p$ as a function of $|t|$ 
for different $\sqrt{s}$ in the TeV energy domain.} 
\label{fi:dsigpi-ptev}
\end{figure}

\begin{figure}[ht]
\epsfxsize=12cm
\epsfysize=16cm
\centerline{\epsfbox{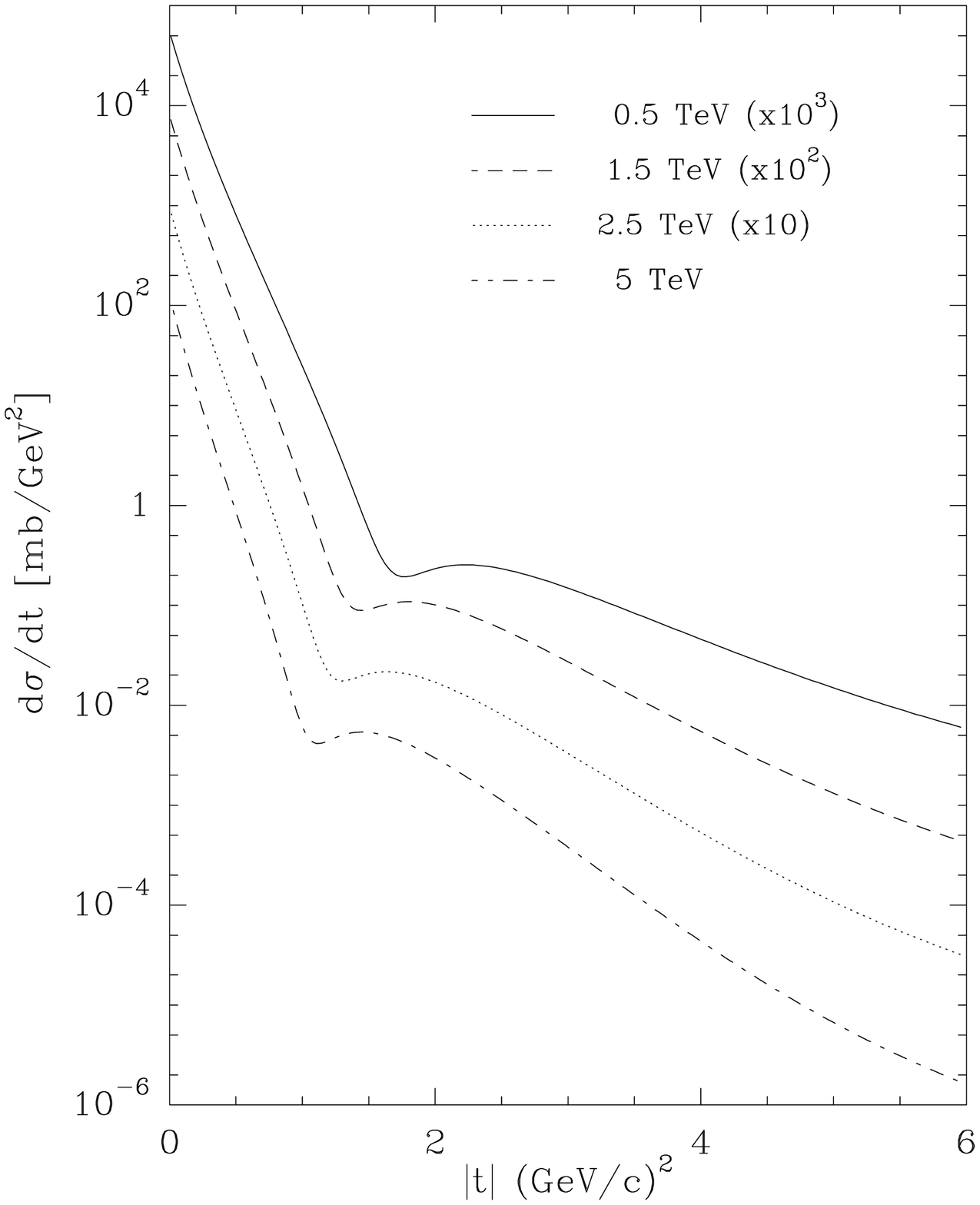}}
\caption{$d\sigma /dt$ for $K^{-} p$ as a function of $|t|$ 
for different $\sqrt{s}$ in the TeV energy domain.} 
\label{fi:dsigk-ptev}
\end{figure}

 The present work is motivated by the realization that the basis 
for the second reason may change in the near future.  The construction of 
the CERN-LHC, which is a proton collider of 7 TeV in each beam, will make 
it possible to have multi-TeV secondary beams of $\pi^{\pm}$ and $K^{\pm}$.  
The scattering of such external secondary beams on protons will provide the 
much needed data for $\pi^{\pm}p$ and $K^{\pm}p$ elastic scattering at 
high energies (see Figs. \ref{fi:dsigpi-ptev} and \ref{fi:dsigk-ptev}). 
For example, the center-of-mass energy of 
a 6 TeV $\pi^{\pm}$ or $K^{\pm}$ and a stationary proton is about 106 GeV, close 
to that of ISR for $pp$ scattering. Encouraged by the likely irrelevance of 
this second reason, it has been possible to overcome the difficulty due to the 
first reason, namely the inclusion of more elaborate Regge backgrounds.

 Looking 
forward, we believe that the present treatment of $\pi^{\pm}p$ and $K^{\pm}p$ 
elastic scattering will make it possible to deal with a number of additional 
scattering processes.  The most interesting process is perhaps the 
following:
\beq{13}
p+p\longrightarrow p + N(1440).
\label{eq:13}
\eeq
This high-energy process was studied nearly 
thirty years ago \cite{r39}.  Since $N(1440)$ has the same $I(J^P)$ as the proton, this 
is one of the simplest  diffraction processes.  This is to be contrasted with the 
elastic scatterings dealt with by most of the existing 
impact-picture phenomenologies, including the present paper. The diffraction 
process (\ref{eq:13}) and many similar ones can perhaps be studied 
experimentally at BNL-RHIC, and we look forward eagerly to such data in the near 
future.\\

{\bf Acknowledgments}

The work of one of us (TTW) was supported in part by the US Department 
of Energy under Grant DE-FG02-84ER40158; he is also grateful for 
hospitality at the CERN Theoretical Physics Division.

%%%%%%%%%%%%%%
%%%%%%%%%%%%%%%%%%%%%%%% P P figures %%%%%%%%%%%%%%%%%%%%%
\newpage
%%%%%%%%
%%%%%%%%%%%%%%%%%
\clearpage
\newpage
%%%%%%%%
%%%%%%%%%%%%%%%%%
\clearpage
%%%%%%%%%%%%%%%%%
\newpage
%%%%%%%%%%%%%%%%%
%%%%%%%%%%%%%%%%%
\clearpage
\newpage
%%%%%%%%%%%%%%%%%
%%%%%%%%%%%%%%%%%
\clearpage
\newpage
%%%%%%%%%%%%%%%%%
%%%%%%%%%%%%%%%%%
\newpage
%%%%%%%%%%%%%%%%%
%%%%%%%%%%%%%%%%%
%%%%%%%%%%%%%%%%%
\clearpage
\newpage
%%%%%%%%%%%%%%%%%
\clearpage
\newpage
%%%%%%%%%%%%%%%%%
\newpage

%%%%%%%%%%%%%%%%%%%%%%%% Pion P figures %%%%%%%%%%%%%%%%%%%%%
%%%%%%%%%%%%%%
\clearpage
\newpage
%%%%%%%%%%%%%%%%%
%%%%%%%%%%%%%%%%%
\clearpage
\newpage
%%%%%%%%%%%%%%%%%
\clearpage
\newpage
%%%%%%%%%%%%%%%%%
%%%%%%%%%%%%%%%%%
\clearpage
\newpage
%%%%%%%%%%%%%%%%%
%%%%%%%%%%%%%%%%%
%%%%%%%%%%%%%%%%%
\clearpage
\newpage
\clearpage
\newpage

\clearpage
%%%%%%%%%%%%%%
%%%%%%%%%%%%%%           KAON-PROTON
\newpage
%%%%%%%%%%%%%%%%%
%%%%%%%%%%%%%%%%%
\clearpage
\newpage
%%%%%%%%%%%%%%%%%
\clearpage
\newpage
%%%%%%%%%%%%%%%%%
%%%%%%%%%%%%%%%%%
\newpage
%%%%%%%%%%%%%%%%%
%%%%%%%%%%%%%%%%%%
\clearpage
\newpage
%%%%%%%%%%%%%%%%%
\clearpage

%%%%%%%%%%%%%%%%%%%%%%%%%%%%%%%%%%%%%%%%%%%%%%%%%%%%%%%%%%%%%%%%%%%%%%%%%%%%%

%%%%%%%%%%%%%%%%%
\end{document}